\documentclass[letter,twocolumn]{aastex702}

\usepackage{upgreek}
\usepackage{bigdelim}
\usepackage{hyperref}

\usepackage{graphicx}
\usepackage{gensymb}
\usepackage{multirow}
\usepackage{amsmath,amssymb,amsfonts}
\usepackage{amsthm}
\usepackage{mathrsfs}
\usepackage[title]{appendix}
\usepackage{xcolor}
\usepackage{textcomp}
\usepackage{booktabs}
\usepackage{listings}

\newcommand{\M}{$M_{\odot}$}
\newcommand{\kms}{km~s$^{-1}$}
\newcommand{\Rvir}{$R_\mathrm{vir}$}
\newcommand{\OI}{$\mathrm{O\ I}$}

\newcommand{\OVI}{$\mathrm{O\ VI}$}
\newcommand{\CI}{$\mathrm{C\ I}$}
\newcommand{\CII}{$\mathrm{C\ II}$}
\newcommand{\CIV}{$\mathrm{C\ IV}$}
\newcommand{\SiII}{$\mathrm{Si\ II}$}
\newcommand{\SiIII}{$\mathrm{Si\ III}$}

\newcommand{\MgII}{$\mathrm{Mg\ II}$}
\newcommand{\NI}{$\mathrm{N\ I}$}
\newcommand{\NII}{$\mathrm{N\ II}$}
\newcommand{\NV}{$\mathrm{N\ V}$}
\newcommand{\CaII}{$\mathrm{Ca\ II}$}
\newcommand{\NaI}{$\mathrm{Na\ I}$}
\newcommand{\SII}{$\mathrm{S\ II}$}
\newcommand{\Lya}{Ly$\alpha$}
\newcommand{\Ha}{H$\alpha$}

\newcommand{\HI}{$\mathrm{H\ I}$}

\newcommand{\HII}{$\mathrm{H\ II}$}
\newcommand{\mH}{H$_{2}$}

\begin{document}

\title{Forming Molecular Hydrogen in a Galaxy Halo Through Star Formation Wind Interactions}
\shorttitle{H$_2$ in CGM}
\shortauthors{Koplitz et al.}

\author[0000-0001-5530-2872]{Brad Koplitz}
\affil{School of Earth \& Space Exploration, 
Arizona State University, 781 Terrace Mall, Tempe, AZ 85287, USA}
\email{brad.koplitz@asu.edu}
\correspondingauthor{Brad Koplitz} 
\email{Brad.Koplitz@asu.edu}

\author[0000-0002-2724-8298]{Sanchayeeta Borthakur}
\affil{School of Earth \& Space Exploration, 
Arizona State University, 781 Terrace Mall, Tempe, AZ 85287, USA}
\email{sborthak@asu.edu}

\author[0000-0001-6670-6370]{Timothy Heckman}
\affil{School of Earth \& Space Exploration, 
Arizona State University, 781 Terrace Mall, Tempe, AZ 85287, USA}
\affil{Department of Physics \& Astronomy, 
Johns Hopkins University, Bloomberg Centre, 
3400 N. Charles Street, Baltimore, MD 21218, USA}
\email{theckma1@jhu.edu}

\author[0000-0003-4237-3553]{Frances H. Cashman}
\affil{Department of Physics, 
Presbyterian College, 
Clinton, SC 29325, USA}
\email{fcashman@presby.edu}

\author[0000-0002-3193-1196]{Evan Scannapieco}
\affil{School of Earth \& Space Exploration, 
Arizona State University, 781 Terrace Mall, Tempe, AZ 85287, USA}
\email{evan.scannapieco@asu.edu}

\author[0000-0002-7982-412X]{Jason Tumlinson}
\affil{Space Telescope Science Institute, 
3700 San Martin Drive, Baltimore, MD 21218, USA}
\affil{Center for Astrophysical Sciences, 
William H. Miller III Department of Physics \& Astronomy, 
Johns Hopkins University, 
3400 N. Charles Street, Baltimore, MD 21218, USA}
\email{tumlinson@stsci.edu}

\author[0000-0001-8813-4182]{Hsiao-Wen Chen}
\affil{Department of Astronomy \& Astrophysics,
University of Chicago, 5640 S. Ellis Avenue, Chicago, IL 60637, USA}
\email{hchen@astro.uchicago.edu}

\author[0000-0003-1455-8788]{Molly Peeples}
\affil{Space Telescope Science Institute, 
3700 San Martin Drive, Baltimore, MD 21218, USA}
\affil{Center for Astrophysical Sciences, 
William H. Miller III Department of Physics \& Astronomy, 
Johns Hopkins University, 
3400 N. Charles Street, Baltimore, MD 21218, USA}
\email{molly@stsci.edu}

\begin{abstract}
Molecular hydrogen ({\mH}) is the key ingredient for star formation in galaxies. 
However, how {\mH} is distributed in and around galaxies is an open question in astronomy. 
We report the detection of {\mH} (covering excitation levels $J = 0$ to 3) in a unique environment at a distance of 139~kpc from a starbursting luminous infrared galaxy. 
This is the furthest from a galaxy that {\mH} has been directly detected to date. 
This {\mH} system is unlike any seen before, with a high molecular-to-atomic hydrogen ratio (2\%) and an accompanying low neutral hydrogen column density (log$_{10}[N(${\HI}$)/\mathrm{cm}^{-2}] = 18.52^{+0.13}_{-3.58}$). 
Our measurements reveal that the {\mH} bearing cloud is traveling at $\sim$100~{\kms} relative to the host galaxy, has a super-solar metallicity, and excites at a single, cold temperature ($245^{+25}_{-21}$~K) indicating a lack of the ``typical'' core-envelope structure seen in other {\mH} clouds.
Furthermore, our analysis points to the {\mH} likely being formed through interactions between galactic winds and the gaseous halo surrounding the galaxy. 
The properties of this cloud suggest that the {\mH} is short-lived, yet it presents crucial information on the microphysics of the wind-CGM interactions.
\end{abstract}


\section{Introduction} 

The processes that cycle gas into and out of galaxies are vital to a complete understanding of how galaxies evolve \citep{Tumlinson2017,PerouxHowk2020,DonahueVoit2022}. 
Feedback from star formation (SF) and active galactic nuclei (AGN) is known to move gas out of a galaxy’s interstellar medium (ISM) and into its halo, known as the circumgalactic medium (CGM), regulating the growth and future SF of a galaxy \citep{Veilleux2020}. 

The flow of molecular hydrogen ({\mH}) is particularly important as it is the necessary fuel for SF with its origin and distribution around galaxies being open questions in astronomy.
While {\mH} forms efficiently on dust grains \citep{Wakelam2017}, theoretical work predicts that it can also arise through rapid cooling behind shocks \citep{Appleton2006,Appleton2017} or by condensation within hot galactic winds associated with outflows \citep{Scannapieco2017,GronkeOh2018,GronkeOh2020,ChenOh2024}.
These mechanisms imply that molecular gas may form \textit{in situ} in galaxy halos, but observational constraints remain scarce.

When observed outside of the Milky Way, {\mH} is typically seen on scales of a few to tens of kpc from the galaxy \citep{Appleton2006,Tchernyshyov2022,YuiDan2025}.
These molecular outflows are typically assumed to be driven by winds generated from SF or AGN \citep{Veilleux2013,Leroy2015,Fluetsch2019,Li2021}.
Starburst galaxies, in particular, are known to have a substantial amounts of {\mH} in their ISM \citep{Hoopes2004} and exhibit outflows driven by stellar feedback; e.g., M82 \citep{Leroy2015,Lopez2025}.

These feedback processes drive galactic winds which are complex environments where hot plasma coexists with warm ionized, neutral, and molecular gas \citep{Guo2023,ThompsonHeckman2024,Ha2025,Veilleux2025}.
As the hot wind fluid sweeps through the CGM, it interacts with cooler clouds within the halo, causing a cascade of heating and cooling processes that is not well understood.
While complex filamentary structures produced by wind-interstellar medium (ISM) interactions have been observed very close to disks of starburst galaxies, such observations beyond a few kpc are not feasible with current technology. 
The {\mH} that is observed further from the disk of galaxies is typically associated with damp or sub-damp Lyman-$\alpha$ absorbers \citep{Richter2003,Noterdaeme2015,Srianand2008,Muzahid2015,Cashman2021}.
Such high amounts of neutral hydrogen allows for self shielding, preventing the destruction of the molecular gas \citep{ChenOh2024}.

Here we report the detection of {\mH} absorption 139~kpc from a starbursting luminous IR galaxy (LIRG). 
This is the furthest that {\mH} has been seen from a galaxy to date. 
This may represent the first-ever detection of complex cooling enabled by wind-CGM interactions far into the halo.
As a result, it may provide a path forward for modeling the physics of the SF-driven feedback impacting the entire galaxy.

The rest of this paper is organized as follows:
Section~\ref{sec:data} details the data and measurements of this study,
Section~\ref{sec:results} discusses the implication of our measurements, 
and Section~\ref{sec:formation} investigates possible origins of the {\mH} gas before summarizing the results in Section~\ref{sec:conclusion}.
Throughout this study we assume a cosmology with $\Omega_m = 0.3$, $\Omega_\Lambda = 0.7$, and $H_0 = 70$~{\kms}\,Mpc$^{-1}$. 

\section{Data and Analysis} \label{sec:data}

\subsection{Galaxy J005712.17+144708.5}

\begin{figure}
    \centering
    \includegraphics[width=0.9\linewidth]{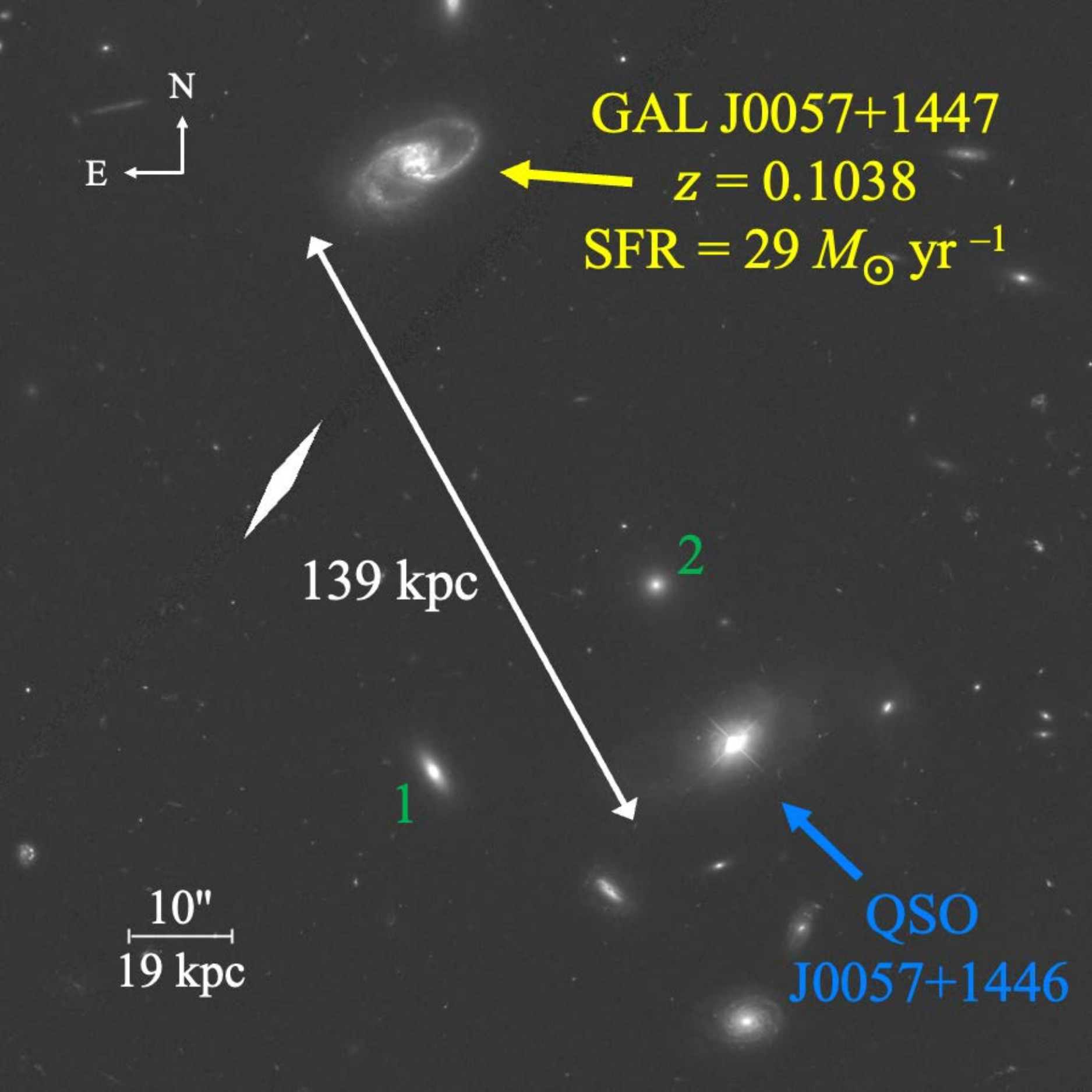}
    \caption{HST image of the QSO $-$ galaxy pair QSO J0057+1446 $-$ GAL J0057+1447. GAL J0057+1447 is indicated in yellow with QSO J0057+1446 in blue. The separation between the QSO and galaxy is shown in white. The two galaxies which are at similar redshifts as GAL J0057+1447 (discussed in Section~\ref{sec:tidal}) are highlighted with numbers. The white diamond on the left side is an artifact of the tiling from HST.}
    \label{fig:image}
\end{figure}

We present the detection of 19 rotational transitions of {\mH} with neutral-atomic hydrogen ({\HI}) and several metal species in absorption in the CGM of the galaxy J005712.17+144708.5, hereafter GAL J0057+1447.
These features were found along the sight line towards the quasi stellar object (QSO) J0057+1446, which pierces the CGM of GAL J0057+1447 at a projected distance of 139~kpc, corresponding to about 11 times the galaxy's $r$-band stellar radius ($R_{\mathrm{pet},r}$) and about half its virial radius ({\Rvir}).
The QSO sight line is positioned along the minor axis of GAL J0057+1447. 
We show an F606W image from the Hubble Space Telescope (HST; PID 10421; PI: G. Canalizo) of the system showing the galaxy, QSO, and environment in Figure~\ref{fig:image}.

The properties of GAL J0057+1447 are presented in Table~\ref{tab:gal_prop}.
We estimated the inclination angle of the galaxy to be 59\degree\ from the Sloan Digital Sky Survey (SDSS) $g$ band image, using the publicly available software \texttt{statmorph} \citep{Rodriguez-Gomez2019}.
The redshift ($z$) and stellar mass ($M_\star$) come from the MPA-JHU catalog for SDSS data release 14 \citep{SDSS_DR14_2018}.
This $z$ points to a luminosity distance to GAL J0057+1447 of 479~Mpc.
The star formation rate (SFR = 29~{\M}~yr$^{-1}$), derived from the IRAS 60~$\mu$m luminosity \citep{Moshir1990,RosaGonzalez2002}, and location on the Baldwin Phillips Terlevich diagram \citep{BPT1981} shows that GAL J0057+1447 is a star-forming galaxy.
The galaxy has a specific SFR (SFR / M$_\star$) of 10$^{-9.6}$~yr$^{-1}$ making it a starburst, similar to the highly studied galaxy M82 ($\gtrsim$10$^{-9.5}$~yr$^{-1}$; \citealt{Dale2023,Yuan2023}) though its $M_\star$ is an order of magnitude higher.
From the large far-IR luminosity relative to the near-ultraviolet luminosity ($L_\mathrm{FIR} \approx 100~L_\mathrm{NUV}$), we conclude GAL J0057+1447 is a luminous infrared galaxy (LIRG).

\begin{deluxetable}{lcl}
    \tablecaption{Galactic Properties of GAL J0057+1447} 
    \label{tab:gal_prop}
    \tablehead{\colhead{Parameter} & \colhead{Value} & \colhead{Units}\\
    \colhead{(1)} & \colhead{(2)} & \colhead{(3)}}
    \startdata
    Alternate ID & 2MASX J00571219+1447089 & \\
    R.A. (J2000) & 00:57:12.17 & \\
    Decl. (J2000) & +14:47:08.6 & \\
    Redshift, $z$ & 0.1038 & \\
    % Distance, $d_\mathrm{L}$ & 479 & Mpc \\
    Inclination, $i$ & 59 & $\degree$ \\
    Star Formation Rate, SFR & 29 & {\M}~yr$^{-1}$ \\
    Stellar Mass, $M_\star$ & 11.1 & log$_{10}(M_\star / M_\odot)$ \\
    Halo Mass, $M_\mathrm{halo}$$^{a}$ & 12.6 & log$_{10}(M_\mathrm{halo} / M_\odot)$ \\
    Virial Radius, {\Rvir}$^{b}$ & 269 & kpc \\
    Stellar $r$-band Radius, $R_{\mathrm{pet},r}$ & 12.8 & kpc \\
    \enddata
    \tablecomments{ $^a$ Calculated using a modified \cite{Kravtsov2018} relation based on the finding of \cite{Mandelbaum2016}. $^b$ Calculated using the \cite{Liang2014} relation.}
\end{deluxetable}

\subsection{Observations} \label{sec:obs}

The UV observations presented here were taken with the Cosmic Origin Spectrograph (COS~\citealt{Osterman2011,Green2012}) aboard HST through proposal GO-17093 (PI:~S.~Borthakur), which targeted QSO J0057+1446 at $z = 0.17209$.
We used the medium-resolution grating G130M ($R \approx 15$,000; FWHM $\approx$ 20~{\kms}) and the low-resolution grating G230L ($R \approx 3$,000; FWHM $\approx$ 100~{\kms}).
This data was reduced using the standard COS pipeline \citep{Rafelski2018}.
The G130M spectrum was binned by three pixels to increase the signal-to-noise of the data and achieve Nyquist sampling.

We also obtained high resolution ($R \approx 38$,000; FWHM $\approx$ 8~{\kms}) optical spectra of the QSO J0057+1446 from the Hectochelle spectrograph \citep{Szentgyorgyi1998} on MMT through PID S200 (2024B Semester; PI: B. Koplitz).
These observations were taken with the OB25 and OB37 filters, allowing us to measure the strength of {\NaI} and {\CaII}, respectively, in the CGM of GAL J0057+1447.
This data was reduced using the standard Hectochelle reduction pipeline \citep{Mink2007}.
To increase the signal-to-noise ratio, we binned the spectra by 2 pixels.

Archival optical spectra and images are available for the QSO J0057+1446, GAL J0057+1447, and the surrounding field.
QSO J0057+1446 and GAL J0057+1447 were observed spectroscopically by SDSS ($R \approx 2$,000; FWHM $\approx$ 150~{\kms}) and the Dark Energy Spectroscopic Instrument (DESI; $R\approx2$,000$-$5,000; FWHM $\approx$ 60$-$150~{\kms}).
The field around QSO J0057+1446 was also observed with MUSE ($R$~$\approx$~3,000; FWHM~$\approx$~100~{\kms}; \citealt{Bacon2010}) on the Very Large Telescope as part of program 097.B-0080 (PI: B. Husemann).
GAL J0057+1447 is not included in the MUSE data-cube; however, it does include galaxies not observed spectroscopically with SDSS.
Optical images of the field around QSO J0057+1446 and GAL J0057+1447 were taken by SDSS, HST and the Dark Energy Camera Legacy Survey (DECaLS).

\subsection{Voigt Profile Analysis} \label{sec:voigt}

To characterize the absorption features of this system, we fit Voigt profiles using a reduced $\chi^2$ algorithm \citep{SembachSavage1992}.
This is the same method employed by the COS-Halos \citep{Tumlinson2013,Werk2013}, COS-GASS \citep{Borthakur2015}, and COS-DIISC \citep{Borthakur2024,Koplitz2025} surveys.
This analysis allows us to determine the column density ($N$), Doppler width ($b$), and velocity relative to GAL J0057+1447 ($v$) of absorbers.
All atomic data used when fitting comes from \cite{Morton2003}.
Before fitting, we normalize the continuum of each absorber within $\pm$600~{\kms} of GAL J0057+1447's systematic velocity ($v_\mathrm{sys}$).
The absorbers beyond $\pm600$~{\kms} are likely not gravitationally bound to the galaxy and so are excluded from our analysis.
Absorbers not associated with GAL J0057+1447 were also identified to evaluate any contamination.

We present our measurements in Appendix~\ref{sec:AppB}.
Figures showing the spectrum and associated Voigt profile fits are available in the online journal.
For transitions not detected at 3$\sigma$ above the noise level, we report an upper limit as the $3\sigma$ error on the rest-frame equivalent width ($W_r$) within $-150$ and $-50$~{\kms} of $v_\mathrm{sys}$, since the detected features are found at $v \approx -100$~{\kms}.
Since the FWHM of the G230L grating, which covers the location of the {\MgII} 2803.5~\AA\ line at the $z$ of GAL J0057+1447, is significantly larger than that of the G130M grating, the 3$\sigma$ upper limit of this transition is calculated over the velocity range from $-200$ to $0$~{\kms}.
These limits are then converted to an upper limit on the the transition's column density assuming the absorber is in the linear regime of the curve of growth. 

When fitting the {\mH} absorbers, we initially assume the Doppler width of each excitation level is independent of the others.
This led to widths of $b$ = $11.5\,\pm\,1.8$, $25.8\,\pm\,1.1$, $15.4\,\pm\,2.5$, and $28.3\,\pm\,3.7$~{\kms} for the $J$ = 0, 1, 2, and 3 level, respectively.
The broader ortho lines ($J$ = 1 and 3) may reflect the environment of the {\mH}. 
At the single, warm temperature we find for the {\mH} ($\sim$244~K, discussed below), the ortho lines are preferentially populated over the para lines ($J$ = 0 and 2) at a ratio of 3:1.
We refit the {\mH} absorbers using a single width of $b = 21.5\,\pm\,1.2$~{\kms} (the weighted mean of the original widths), similar to other studies \citep{Cashman2021,Tchernyshyov2022}.
Both sets of Voigt profile fits lead to similar results.

The {\CIV} 1550.8~\AA, {\NV} 1238.8~\AA, and {\OVI} 1037.6~\AA\ lines are blended with intervening systems. 
For the {\NV} line, we calculate the 3$\sigma$ upper limit over the velocity range $\pm$50~{\kms}.
No measurement is able to be obtained for the {\CIV} or {\OVI} lines since the blending is too severe. 

It is important to note that measurements derived from Voigt profile fitting have limitations.
The resolution of COS can cause multiple, narrow clouds at similar velocities to appear as a single, wider component.
Hints of this can be seen in the high-resolution optical spectra from Hectochelle which shows multiple {\NaI} components.
We conclude from this that the UV absorbers, particularly the {\mH} which is expected to have a thermal width of a few {\kms}, are likely also being produced by multiple components.
We attempted to refit the UV absorbers while fixing the velocity centroids to those of {\NaI}; however, the signal-to-noise ratio and resolution of the COS spectra prevented these fits from converging. 
As a result, we assume that the measurements presented here are dominated by the largest absorbing cloud, though a more complicated picture may be revealed by higher-resolution observations.

\section{Results} \label{sec:results}

\subsection{Properties of the Hydrogen Absorption}

The detected {\mH} covers four excitation levels ($J = 0, 1, 2,$ and 3) with an estimated column density ($N$) of log$_{10}[N(${\mH}$)/\mathrm{cm}^{-2}] = 16.52\,\pm\,0.05$.
We also detect atomic hydrogen via the Lyman-$\alpha$ ({\Lya}) transition and measure a column density of log$_{10}[N(${\HI}$)/\mathrm{cm}^{-2}] = 18.52^{+0.13}_{-3.58}$.
The {\Lya} feature is saturated and does not show damping wings, discussed more in Appendix~\ref{sec:HI}.
This enables us to confidently limit the upper bound for $N$({\HI}); however, the saturation leads to an uncertain lower limit.
While the $N$({\mH}) we measure is consistent with ISM-like values \citep{Tumlinson2002,Gillmon2006,Cashman2021,Tchernyshyov2022}, the $N$({\HI}) of this system is significantly smaller.

We measure the {\mH} fraction of the system to be

\begin{equation}
f(\mathrm{H}_2) = \frac{2N(\mathrm{H}_2)}{2N(\mathrm{H}_2)\ +\ N(\mathrm{H\ I})} = 0.02^{+0.94}_{-0.01}.
\end{equation}

\begin{figure*}
    \centering
    \includegraphics[width=\linewidth]{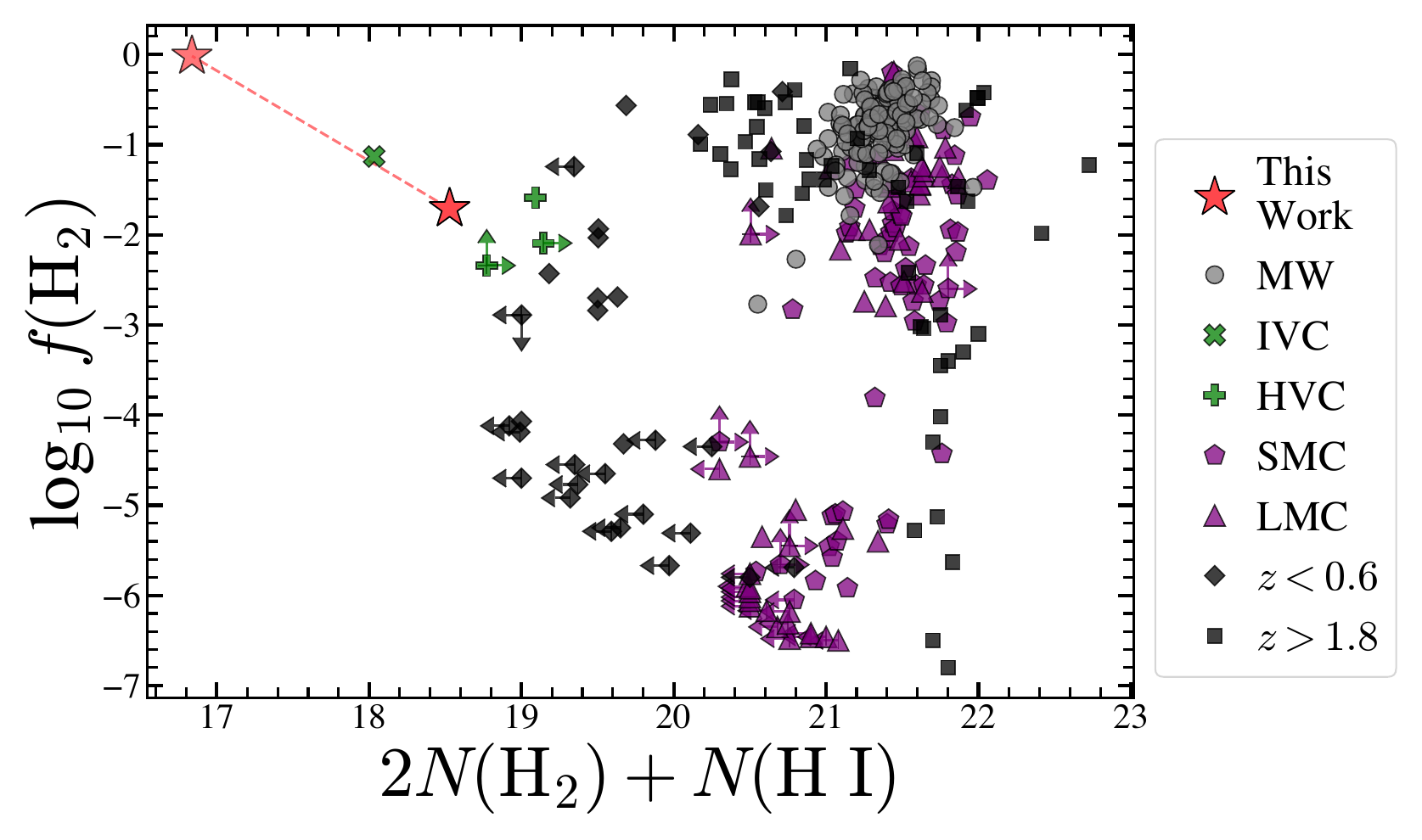}
    \caption{Molecular hydrogen fraction as a function of the neutral hydrogen column density. Both {\HI} measurements are shown and connected by a dashed line to indicate the possible parameter space this system could occupy. For comparison, we also plot sources from the Milky Way (grey circles; \citealt{Rachford2009,Shull2021}), Intermediate Velocity Clouds (green x; \citealt{Richter2003}), High Velocity Clouds (green +; \citealt{Cashman2021,Tchernyshyov2022}), the Small and Large Magellanic Clouds (purple pentagons and triangles, respectively; \citealt{Welty2012}), as well as sub-damped and damped {\Lya} systems at $z < 0.6$ (black diamonds; \citealt{Crighton2013,Muzahid2015,Muzahid2016,Boettcher2021,Szakacs2021}) and $z > 1.8$ (black squares; \citealt{Srianand2008,Prochaska2009,Balashev2014,Balashev2019,Ranjan2020}).}
    \label{fig:H2frac}
\end{figure*}

This system lands in a unique part of the $f$({\mH}) and neutral H column density parameter space (Figure~\ref{fig:H2frac}).
We compare our measurements to those from the literature of ISM systems across redshift in Figure~\ref{fig:H2frac}, showing that this absorbing system is not ISM-like.
We include the uncertainty in $N$({\HI}) in the figure as a dashed line.
Only one literature source comes close to this system, but the $N$({\HI}) for that target was inferred from $N$({\OI}) rather than directly measured \citep{Richter2003}.
This makes our system the lowest neutral H column density cloud which bares {\mH}.

\begin{figure}[ht]
    \centering
    \includegraphics[width=\linewidth]{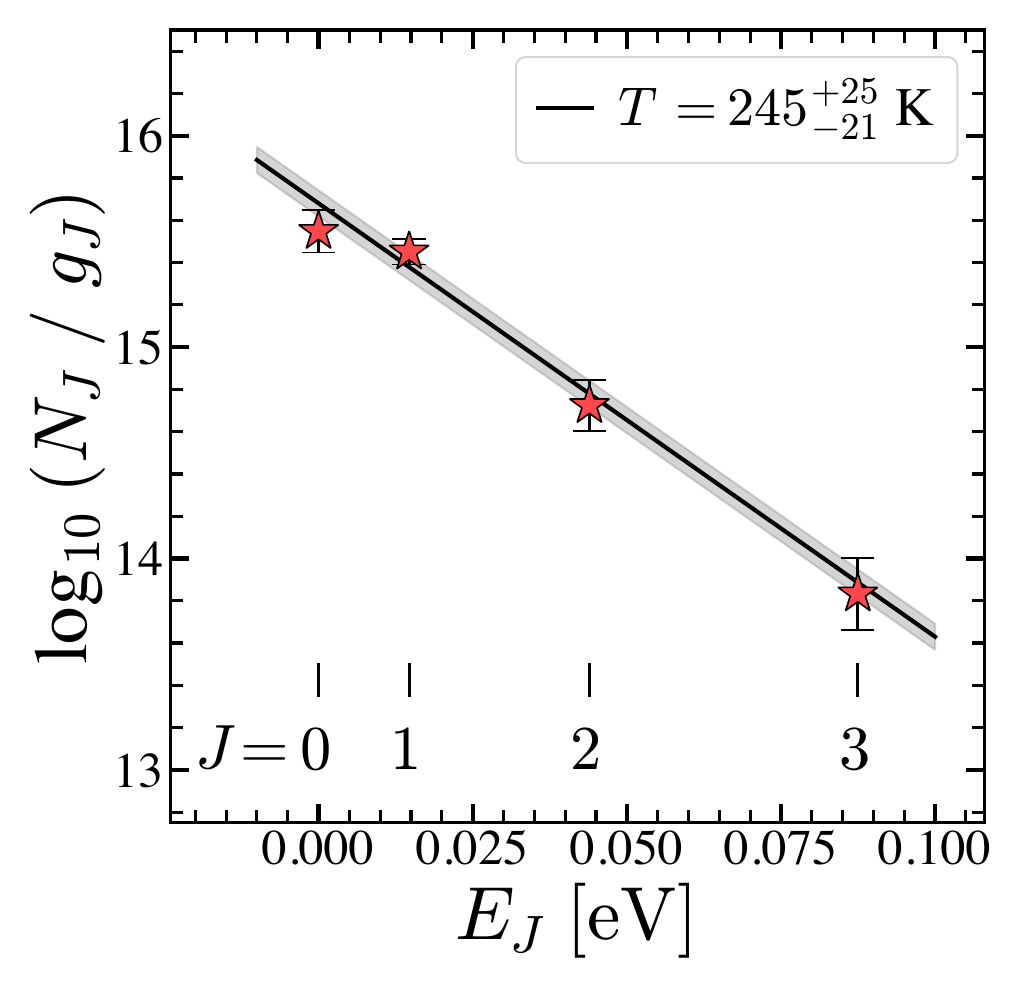}
    \caption{Excitation diagram of {\mH}. The energy of each excitation level ($J$) is indicated at the bottom of the figure. The line of best fit is shown in black, with the corresponding temperature shown in the legend. The uncertainty in the temperature is shown as a gray shaded region.}
    \label{fig:excitation}
\end{figure}

The temperature of the {\mH} cloud can be constrained using the measurements of multiple excitation levels, known as the excitation diagram.
We show the excitation diagram of our system in Figure~\ref{fig:excitation}.
This indicates that the {\mH} in the CGM of GAL J0057+1447 is consistent with a single, cold temperature of $T = 245^{+25}_{-21}$~K.
{\mH} is typically found to be fit by a multi-temperature solution, with the ground states ($J = 0$ and 1) being fit by a colder temperature than the more excited states ($J \geq 2$; \citealt{Richter2003,Shull2021,Cashman2021,Tchernyshyov2022}).
This type of solution is indicative of a cold core encased within a warmer shell.
It is unique to find {\mH} without a core-envelop geometry and may hint at the origin of the gas.
Fitting our measurements in this way gives $T = 762$~K for the ground state transitions and $T = 246$~K for the excited states. 
This sort of temperature inversion has not been seen before and would be puzzling to explain.
Additionally, the Bayesian Information Criterion (BIC) is roughly equal for both models (BIC = 6.26 and 6.94 for the single and two temperature solutions, respectively).
These lead us to conclude that the {\mH} gas is likely at a single, cold temperature.

\subsection{Properties of the Metal Absorption}

We also find absorption features from 12 metal species tracing a range of gas phases; the neutral phase from {\NaI}, {\CI}, {\OI}, and {\NI}, the low-ionization phase from {\MgII}, {\SiII}, {\CII}, {\NII}, and {\SiIII}, and the high-ionization phase from {\CIV}, {\NV}, and {\OVI}.
Most of the ionized metals ({\CII}, {\SiIII}, {\CIV}, {\NV}, and {\OVI}) are consistent with measurements from other CGM surveys \citep{Werk2013,Borthakur2016,Koplitz2025} and are likely associated with the ambient CGM.

We estimate the metallicity of the absorbing cloud using multiple methods.
Since {\OI} and {\HI} are locked together by charge exchange \citep{FieldSteigman1971} and oxygen is not significantly depleted onto dust \citep{Jenkins2009}, we can assume that {\OI}\,/\,{\HI} = O\,/\,H.
From this, we estimate the metallicity to be $\gtrsim$1.9~$Z_\odot$.
It is important to note that the {\OI} absorption is saturated, which makes this value a lower limit.
Using the upper limit of {\SII}, we obtain an upper limit on the metallicity of $<$5.2~$Z_\odot$.
These show that our system has at least solar metallicity, and likely super-solar, when the CGM is typically found to have a sub-solar metallicity \citep{Lehner2013,Prochaska2017}.
We note that outflowing winds, in particular, have been found over a wide range of metallicities ($<$20\% up to $\sim$320\% solar; \citealt{Ashley2022}).

\subsection{Ionization Modeling}

To determine if these absorbers are in photoionization equilibrium (PIE), we compare our measurements to \texttt{CLOUDY} models (v.23;~\citealt{Chatzikos2023}). 
We generated models with metallicities of $Z =$ 1, 2, 3, and 5~$Z_\odot$ and solar relative abundances \citep{Asplund2009}.
The total hydrogen volume density was allowed to vary from log$_{10}[n(\mathrm{H})/\mathrm{cm}^{-3}] = -6.00$ to $4.00$ in steps of $0.25$~dex.
Due to the lack of multiple Lyman lines, we also vary $N$({\HI}) between log$_{10}[N(${\HI}$)/\mathrm{cm}^{-2}] = 15.00$ and $18.75$ in steps of 0.25~dex.
This range is set based on the best-fit {\HI} models found for the {\Lya} absorption feature.
Each model was exposed to the extragalactic UV background (EUVB; \citealt{HM2012}).
We exclude {\CaII} and {\SII} from this analysis since they were not detected at 3$\sigma$.
We also exclude {\CII}, {\SiIII}, {\CIV}, {\NV}, and {\OVI} since these absorbers are likely not associated with our {\mH} cloud, as discussed above.

To replicate our measurements in Figure~\ref{fig:H2frac}, the PIE models required log$_{10}[N(\mathrm{H\ I})/\mathrm{cm}^{-2}] = 18.50$ or $18.75$, log$_{10}[n(\mathrm{H})/\mathrm{cm}^{-3}] = 2.50 - 3.00$, and $Z = 1$ or $2~Z_\odot$.
We are able to estimate the cloud length from these models as $L$ = $N$(H) / $n$(H), where we assume $N$(H) = $N$({\HI}) + $N$({\HII}) + 2$N$({\mH}), though $N$({\HII}) is negligible in a system such as this.
These models imply cloud lengths of approximately $10^{-3}$ and $10^{-2}$~pc or about 200 to 2000~AU.
Considering only the metal absorbers, we are unable to reproduce all of the measurements simultaneously.
A single model was able to match, at best, 4 of the 7 metal species, though they always underproduce the {\mH} by over 2 orders of magnitude.
These indicate that the system of absorbers in the CGM of GAL J0057+1447 is inconsistent with a single-phase PIE model.

To determine if these absorbers are collisionally ionized, we compare our measurements to models from \cite{GnatSternberg2007}, which assume solar relative abundances \citep{Asplund2009}.
Since these models give fractional abundances as a function of temperature, we look at the ratio of metals rather than at $N$ directly. 
None of the models (equilibrium, nonequilibrium isobaric cooling, nor nonequilibrium isochoric cooling) were able to reproduce our measurements, regardless of the metallicity.
From these we are able to conclude that the ionization we see in this absorbing cloud is not driven by collisions.

\section{Investigating Formation Scenarios} \label{sec:formation}

Understanding the origin of this system has implications for the role of galactic winds in impacting the CGM.
Below we discuss possible formation scenarios including the QSO sight line probing the ISM of a dwarf galaxy hiding along the QSO sight line, debris from previous tidal interactions, molecular outflows, and \textit{in situ} formation during wind-CGM interactions.
Ultimately, we concluded that the most likely scenario is the {\mH} forming \textit{in situ} during wind-CGM interactions.

\subsection{Foreground Dwarf Galaxy Formation Scenario Analysis}

We consider if a dwarf galaxy could be located along the QSO sight line, causing the {\mH} absorption.
A dwarf galaxy may not be detected in images due to the instrument's large point spread function, making the vicinity of the QSO unsuitable for a photometric search.
Therefore, we investigate if there is any sign of ongoing SF, which is expected for a molecular gas-rich ISM.
No {\Ha} emission is seen in the SDSS spectrum of QSO J0057+1446, indicating that little to no ongoing SF is taking place at the projected position of the QSO. 

We determine the SFR a galaxy could have and be hidden within the spectrum to be 0.01~{\M}\,yr$^{-1}$ \citep{KennicuttEvans2012}.
While this SFR is above that of the Small Magellanic Cloud ($\sim$0.001~{\M}\,yr$^{-1}$; \citealt{Hagen2017}), the super-solar metallicity we infer conflicts with the sub-solar metallicity seen in such low-mass galaxies \citep{Lee2005,Choudhury2020}.
Additionally, the $f$({\mH}) of our system is very different from the SMC, see Figure~\ref{fig:H2frac}.
These lead us to conclude that there is not a foreground dwarf galaxy along the line of sight.

\subsection{Tidal Interaction Formation Scenario Analysis} \label{sec:tidal}

Another possible source of {\mH} is tidal streams and debris.
In particular, {\mH} absorption has been observed in the intragroup medium \citep{Muzahid2015,Boettcher2021,Beckett2023}.
Archival SDSS data shows that GAL J0057+1447 is likely a member of a galaxy group with 4 other massive ($M_\star > 10^{10.9}$~{\M}) members (Group Number 15861; \citealt{Tinker2020,Tinker2021}).
GAL J0057+1447 and QSO J0057+1446 are 221 and 360~kpc from the group center, respectively.
The optical and radio morphology of GAL J0057+1447 does not show any sign of ram pressure stripping \citep{SDSS_DR7_2009,Shimwell2026}.
Additionally, the $f$({\mH}) we measure in disagreement with stripped ISM, as seen in Figure~\ref{fig:H2frac}.
Finally, the intragroup medium has been found to struggle to induce ram pressure stripping in Hickson compact groups \citep{Rasmussen2008}, suggesting that it is unlikely to be occurring in a looser group such as the one GAL J0057+1447 is in.
While these cannot definitely rule out interactions between GAL J0057+1447 and the wider intragroup medium, we conclude that these are not likely to be producing the observed {\mH}. 

Archival MUSE data indicates that GAL J0057+1447 has two satellite galaxies 62 and 36~kpc from the QSO sight line, which are labeled in the HST image in Figure~\ref{fig:image}.
Using the $g$ and $r$ magnitudes measured by SDSS, we estimate their stellar masses to be $M_\star \approx 10^{9.9}$ and $10^{9.4}$~{\M} (6\% and 1\% of GAL J0057+1447, respectively; \citealt{Bell2003}).
The optical morphology of each satellite appears to be symmetric, indicating that major tidal interactions are not taking place in this system.
These suggest that the {\mH} is likely not the result of interactions between GAL J0057+1447 and these satellite galaxies; however, we cannot definitely rule this out.

\subsection{Stellar Outflow Formation Scenario}

Given that the absorber is 139~kpc from a LIRG with a high SFR, one major possibility is the {\mH} absorbers is related to SF-driven feedback \citep{ThompsonHeckman2024}. 
These winds can transport cool ISM, including molecular gas, to large radii away from the galactic disk \citep{Veilleux2020}. 
The super-solar metallicity that we infer is consistent with the wide range of values seen in the Milky Way's winds and the disk of other galaxies \citep{Chisholm2018,Ashley2022}. 
However, the single temperature we find the {\mH} at is in conflict with the core-envelope structure seen in outflows \citep{Cashman2021,Tchernyshyov2022}.

The velocities and timescales required for SF-driven outflows to have transported the {\mH} to where we observe it makes this scenario unlikely.
Assuming lifetimes for molecular outflows between 0.3 and 20~Myr \citep{ChenOh2024,Fisher2025}, outflows from GAL J0057+1447 would need to have velocities of $\gtrsim$6,800~{\kms} to travel 139~kpc.
This is significantly larger than what has been seen for molecular outflows ($\sim$200$-$300~{\kms}; \citealt{Leroy2015,Krieger2021}).
At these typical velocities, the molecular clouds would need to survive for $\gtrsim$450~Myr, which is extremely unlikely.

A similar argument shows that outflows from the satellite galaxies are also unlikely to be responsible for the {\mH}.
The MUSE spectra of the satellite galaxies shows they do not contain ongoing SF, indicating they do not currently host outflows; however, it is possible that they hosted outflows in the past.
To travel 62 and 36~kpc, the outflows would need to have velocities or $\gtrsim$3,000 and $\gtrsim$1,800~{\kms}, respectively, which is again much larger than what is seen in molecular outflows.
At more typical velocities, it would still take $\gtrsim$100~Myr for any molecular gas to reach the QSO sight line. 
This is, again, an unlikely timescale for the {\mH} to have survived.

Simulations have revealed that cold clouds can actually grow in a wind-driven medium, particularly if the {\HI} is self shielding \citep{ChenOh2024}.
The {\HI} measurement of this system makes this unlikely, though.
The rarity of this system suggests that cold outflows are less probable so far out into the halo.

Although the $f$({\mH}) of the system is nearest to the MW HVCs in Figure~\ref{fig:H2frac}, these are found at distances $\lesssim$50~kpc from the disk and typically have metallicities that are near or below solar \citep{Collins2007,Wakker2008,Lehner2009,Richter2015,Fox2016}.
This differs from our system which is located 2 to 3 times further and likely has a super-solar metallicity. 
Additionally, the timescale for HVCs to recycle is $\sim$Gyr while molecular clouds survival times are $\sim$10s of Myr.
Taken together, these show that it is difficult for the {\mH} to have originated from stellar outflows or an HVC-like cloud, though we are not able to definitively rule out this scenario.

\subsection{\textit{In Situ} Formation Scenario Analysis}

SF-driven winds from galaxies are known to interact with the CGM \citep{Borthakur2013,Heckman2017}.
It is possible that such winds can trigger \textit{in situ} {\mH} production via cooling \citep{Appleton2006,Appleton2017,Scannapieco2017} or turbulence in the shocked intragroup medium \citep{Guillard2012}.
The presence of dust in the CGM will further enhance the molecule formation process \citep{Wakelam2017}. 

Here we developed a simplified model of the reactions taking place during shock-induced cooling to determine whether they could produce the {\mH} we observe.
The reactions we account for in our modeling, and their rate coefficients, come from \cite{GloverAbel2008}.
All two body reactions that do not contain deuterium are included, assuming Case B recombination where applicable.
We account for cooling through hydrogen and helium \citep{DalgarnoMcCray1972,Black1981}, {\mH} \citep{GalliPalla1998}, and metals \citep{SutherlandDopita1993}.
We also include the ionization, photodisassociation, and heating rates by a EUVB \citep{HM2012}.

We initialize the model to be full-ionized at $T = 10^5$~K, $n$(H) = $10^{1}$~cm$^{-3}$, and $Z = 0.3~Z_\odot$ and then cool the gas such that it maintains a constant density.
From this we are able to match our measured $f$({\mH}), leading us to conclude that this is a viable formation scenario for the {\mH} we observe in the CGM of GAL J0057+1447.
However, it is important to note that the {\mH} is likely short-lived as heating and ionization processes in the CGM would be effective for this optically thin cloud.
We also note that we do not perform a full exploration of the possible conditions which can reproduce the $f$({\mH}), but rather show that it is possible. 
More detailed modeling is required to accomplish this.

We stress that while our simplified model shows that sufficient {\mH} can form through nonequilibrium processes in the absence of dust, the presence of even trace levels of dust will greatly enhance molecule formation relative to this baseline. 
While current observations do not allow us to determine if there is dust in the CGM of GAL J0057+1447, the super-solar metallicity and high abundance of neutral gas-phase metals suggest that this is likely.
We again note that this modeling only attempts to match the measured $f$({\mH}).

This formation scenario is further supported by the {\mH} being at a single, cold temperature, rather than the multi-temperatures seen in outflows \citep{Cashman2021,Tchernyshyov2022}.
As a result, we conclude that the most likely scenario based on currently available data is wind-CGM interactions causing the gas to compress and cool, forming {\mH} in the process, though we cannot definitively rule out all other possible formation mechanisms.

\subsection{Other Formation Scenarios}

More exotic possibilities for creating and sustaining cold gas exist in the literature, but our understanding of the conditions necessary for these processes to be effective in the CGM is limited and require strong assumptions.
For example, the cold gas may also be encased in hotter material, preventing its destruction as it travels \citep{Fox2004}.
Additionally, magnetic fields could suppress thermal conduction in the cloud, allowing the gas to survive in the CGM \citep{BruggenScannapieco2016}.
Similarly, cosmic rays are hypothesized to transport cold, dense gas to large distances \citep{Hopkins2021}.
While we cannot rule out these scenarios, they have similar limitations as wind-driven outflows.
As a result, we conclude that \textit{in situ} formation via wind-CGM interactions is the most plausible origin supported by our evidence.

\section{Summary} \label{sec:conclusion}

Here we report the detection of 19 absorption features from rotational transitions of {\mH}, alongside {\HI} and 12 metal species, in the CGM of GAL J0057+1447.
These are found 139~kpc from the galaxy, corresponding to approximately 11 $R_{\mathrm{pet},r}$ or half {\Rvir}.
This is the first time that {\mH} gas has been directly detected this far away from the host galaxy.
GAL J0057+1447 is likely a LIRG undergoing a starburst and hosts large SF-driven galactic winds.

We find that interactions between the SF-driven winds and the CGM caused gas in the CGM to cool, creating the {\mH} we observe.
This scenario is supported by multiple pieces of evidence.
The metal absorption measurements point to the system having a super-solar metallicity.
This cloud resides in a different part of the parameter space of Figure~\ref{fig:H2frac} from literature sources of the ISM across redshift, showing this is a unique system.
The single, cold temperature of $245^{+25}_{-21}$~K that we find for the {\mH} shows it likely is being produced through a different mechanism than the multi-temperature clouds found in outflows.
Finally, modeling of the reactions taking place after these wind-CGM interactions showed that we can reproduce the cloud via cooling, even without dust.
This example provides the first evidence that cooling occurs in the CGM of starburst galaxies that can produce extremely cold gas.
The expected short lifetime implies a low covering fraction which may have prevented it from being observed until now.

While this cloud provides insights into the physics taking place in the CGM, particularly as it interacts with SF-driven winds, we are limited by the lack of similar systems which have been observed. 
More examples are needed to conclusively show that these interactions are ubiquitous in starbursting galaxies; however, this is difficult to achieve with today's instruments.
Next generation missions such as the Habitable Worlds Observatory and Extremely Large Telescope will be capable of making great strides in overcoming this limitation.
Through these, more examples of {\mH} forming in the hot CGM can be identified which will allow us to gain insights into our cosmic origins through the processes taking place in galactic ecosystems.

\begin{acknowledgments}
We thank the anonymous reviewer for there helpful comments which greatly improved the clarity of the manuscript.
B.K. thanks Edward Buie II, Yakov Faerman, Jessica K. Werk, Matthew McQuinn, Jonathan Stern, and Cassandra Lochhaas for the fruitful conversations regarding this system.
B.K. and S.B. thank Zhijie Qu for helping identify the archival MUSE observations of QSO J0057+1446 and the surrounding field. 
B.K. and S.B. were supported by NASA/STScI through grant HST-GO-17093.
B.K., S.B., and E.S. acknowledge the native people and the land that Arizona State University’s campuses are located in the Salt River Valley.
Some of the data presented in this article were obtained from the Mikulski Archive for Space Telescopes (MAST) at the Space Telescope Science Institute. 
The specific observations analyzed can be accessed via \dataset[doi:10.17909/vehe-4d78]{https://doi.org/10.17909/vehe-4d78}.
The ancestral territories of Indigenous peoples, including the Akimel O’odham (Pima) and Pee Posh (Maricopa) Indian Communities, whose care and keeping of these lands allows us to be here today.
\end{acknowledgments}

\begin{contribution}
B.K. organized and wrote the manuscript.
S.B. led the HST observing proposal with T.H., J.T., B.K., and E.S. contributing to its development.
B.K. and S.B. led the MMT observing proposal.
B.K. performed the measurements and subsequent analysis of the HST and MMT spectra.
H.W.C. identified the environment that GAL J0057+1447 resides in.
S.B., F.H.C., and E.S. assisted with the analysis. 
S.B., T.H., F.H.C., E.S., and H.W.C. assisted in the interpretation of the results.
All authors provided feedback on the manuscript.
\end{contribution}

\facilities{HST(COS); MAST; Sloan.; MMT (Hectochelle); VLT (MUSE)}

\software{\texttt{matplotlib} (v3.2.2;~\citealt{Hunter2007}), \texttt{astropy} (v6.0.1;~\citealt{Astropy2013,Astropy2018,Astropy2022}), \texttt{numpy} (v1.22.0;~\citealt{Harris2020}), \texttt{scipy} (v1.6.2;~\citealt{Virtanen2020}), \texttt{pandas} (v1.3.5;~\citealt{Reback2021}), \texttt{CLOUDY} (v.23;~\citealt{Chatzikos2023}).}

\appendix

\section{Uncertain {\HI} Measurement} \label{sec:HI}

We measure the amount of atomic hydrogen in this system via the {\Lya} transition in absorption, which is saturated and does not show damping wings.
This led to two Voigt profiles fitting the profile.
One model with log$_{10}[N(${\HI}$)/\mathrm{cm}^{-2}] = 18.52 \pm 0.13$ and $b$ = $40.6 \pm 1.1$~{\kms} while the other had log$_{10}[N(${\HI}$)/\mathrm{cm}^{-2}] = 15.44 \pm 0.50$ and $b$ = $82.8 \pm 1.2$~{\kms}.
The model with larger $N$({\HI}) is statistically-preferred ($\chi^2_\mathrm{DOF} = 0.4914$) over the model with smaller $N$({\HI}) ($\chi^2_\mathrm{DOF} = 0.5603$).
We also compare the {\Lya} absorption to models with $N$(\HI) = $10^{19.0}$, $10^{19.5}$, and $10^{20.0}$~cm$^{-2}$ with $b$ = 5~{\kms}.
These stronger profiles are not statistically preferred to the best-fit models presented above and shows that there are no damping wings in the {\Lya} spectrum.

Since our analysis found a bimodal distribution in the best-fit Voigt profiles, we explore the consequences of adopting the smaller {\HI} measurement ($N$({\HI}) = $10^{15.44}$~cm$^{-2}$).
This leads to $2N(\mathrm{H}_2)\ +\ N(\mathrm{H\ I})$ = $10^{16.84}$~cm$^{-2}$ and $f$({\mH}) = 0.96 or 96\%.
These would place our system in Figure~\ref{fig:H2frac} as the highest $f$({\mH}) value by $>$20\% but with a corresponding neutral hydrogen column density that is nearly 5 orders of magnitude smaller than the system with the next highest $f$({\mH}).
{\HI} gas has been shown to provide self shielding from ionizing photons when $N$({\HI}) $\gtrsim$ $10^{17}$~cm$^{-2}$ \citep{ChenOh2024}; however, this $N$({\HI}) is substantially smaller, meaning the {\mH} would likely have been destroyed by the extragalactic UV background \citep{HM2012}.
Comparing this measurement to PIE models, \texttt{CLOUDY} under predicts the amount of metals by at least 2 orders of magnitude.
Additionally, the {\OI}\,/\,{\HI} ratio suggests a metallicity over 2300~$Z_\odot$, which is not physical, though this is impacted by the saturated {\OI} profile.
Because of these, we adopt the statistically preferred, larger $N$({\HI}) measurement for the results presented in this manuscript, though we include the smaller $N$({\HI}) in the reported uncertainties since we cannot definitively excluding the smaller $N$({\HI}) model.
It is important to note again that the absorber is likely comprised of multiple, overlapping components which the COS data is unable to resolve and that the these models should be thought of as limits rather than a robust measure of $N$({\HI}).

\section{Absorption Line Measurements} \label{sec:AppB}

Here we present the measurements from our Voigt profile analysis.
The fitting process is discussed in Section~\ref{sec:voigt}.

\startlongtable
\begin{deluxetable}{lccccc}
    \label{tab:metal_measurements} 
    \centering 
    \tablecaption{Absorption Line Measurements} 
    \tablehead{\colhead{Species} & \colhead{$\lambda_\mathrm{rest}$ $\left[\mathrm{\AA}\right]$} & \colhead{$W_r$ $\left[\mathrm{m\AA}\right]$ $^{a}$} & \colhead{log$_{10}\left(N / \mathrm{cm}^{-2}\right)$ $^{b}$} & \colhead{$b$ $\left[\mathrm{km\ s}^{-1} \right]$ $^c$} & \colhead{$v$ $\left[\mathrm{km\ s}^{-1} \right]$}}
    \startdata
     H$_2$ $J = 0$ &  &  & $15.55\pm0.10$ & $21.5\pm1.2$ & $-$93.9 $\pm$ 5.1 \\
     $0 - 0$, R(0) & 1108.1 & 65.2 $\pm$ 20.6 &  &  \\
     $1 - 0$, R(0) & 1092.2 & 73.3 $\pm$ 24.3 &  &  \\
     $2 - 0$, R(0) & 1077.1 & 114.3 $\pm$ 17.6 &  &  \\
     $3 - 0$, R(0) & 1062.9 & 142.7 $\pm$ 20.2 &  &  \\
     $4 - 0$, R(0) & 1049.4 & 153.9 $\pm$ 25.6 &  &  \\
     \hline
     H$_2$ $J = 1$ &  &  & $16.41\pm0.06$ & $21.5\pm1.2$ & $-$92.8 $\pm$ 2.3 \\
     $0 - 0$, P(1) & 1110.1 & 93.9 $\pm$ 16.7 &  &  \\
     $0 - 0$, R(1) & 1108.6 & 91.0 $\pm$ 18.5 &  &  \\
     $1 - 0$, P(1) & 1094.1 & 98.1 $\pm$ 15.8 &  &  \\
     $1 - 0$, R(1) & 1092.7 & 127.5 $\pm$ 17.4 &  &  \\
     $3 - 0$, P(1) & 1064.6 & 192.2 $\pm$ 19.3 &  &  \\
     $3 - 0$, R(1) & 1063.5 & 222.3 $\pm$ 17.1 &  &  \\
     $4 - 0$, R(1) & 1050.0 & 199.1 $\pm$ 22.4 &  &  \\
     \hline
     H$_2$ $J = 2$ &  &  & $15.42\pm0.12$ & $21.5\pm1.2$ & $-$96.0 $\pm$ 6.1 \\
     $2 - 0$, R(2) & 1079.2 & 97.9 $\pm$ 16.6 &  & \\
     $3 - 0$, P(2) & 1066.9 & 84.2 $\pm$ 20.2 &  & \\
     $3 - 0$, R(2) & 1065.0 & 128.2 $\pm$ 20.2 &  & \\
     $4 - 0$, P(2) & 1053.3 & 99.2 $\pm$ 22.8 &  & \\
     \hline
     H$_2$ $J = 3$ &  &  & $15.16\pm0.17$ & $21.5\pm1.2$ & $-$107.9 $\pm$ 11.9 \\
     $3 - 0$, P(3) & 1070.1 & $<$68.7&  &  \\
     $3 - 0$, R(3) & 1067.5 & 63.1 $\pm$ 19.7 &  &  \\
     $4 - 0$, R(3) & 1054.0 & 85.7 $\pm$ 27.0 &  &  \\
     $5 - 0$, P(3) & 1043.5 & 89.9 $\pm$ 29.8 &  &  \\
     \hline
     Total H$_2$ &  &  & 16.52 $\pm$ 0.05 & \\
     \hline
     {\HI} & 1215.7 & $1109.0\pm43.2$ & $18.52^{+0.13}_{-3.58}$ & $40.6^{+43.4}_{-1.1}$ & $-110.9^{+5.8}_{-7.7}$ \\
     \hline
     {\NaI} & 5891.6 & 394.3 $\pm$ 15.5 & 11.66 $\pm$ 0.04 & 5.34 $\pm$ 1.2 & $-$111.4 $\pm$ 0.5 \\
     {\NaI} & 5897.6 & 320.9 $\pm$ 14.4 & 11.91 $\pm$ 0.21 & 1.03 $\pm$ 1.4 & $-$92.3 $\pm$ 0.5 \\
      & & & 12.00 $\pm$ 0.02 & 11.9 $\pm$ 1.1 & $-$72.0 $\pm$ 0.5 \\
     \hline
     {\CI}& 1280.1 & 94.5 $\pm$ 17.2 & 14.77 $\pm$ 0.11 & 19.0 $\pm$  1.8 & $-$107.4 $\pm$ 4.8 \\
     \hline
     {\CaII} & 3934.8 & $<$98.1 & $<$12.05 & ... & ... \\
     {\CaII} & 3969.6 & $<$67.8 & & \\
     \hline
     {\OI} & 1039.2 & 172.6 $\pm$ 35.3 & 15.54 $\pm$ 0.12 & 31.7 $\pm$ 10.6 & $-$103.5 $\pm$ 6.8 \\
     \hline
     {\NI} & 1134.2 & 84.0 $\pm$ 22.2 &  &  \\
     {\NI} & 1134.4 & 128.2 $\pm$ 22.2 &  &  \\
     {\NI} & 1135.0 & 168.7 $\pm$ 22.3 &  \multirow{2}{*}{14.92 $\pm$ 0.07} &  \multirow{2}{*}{22.6 $\pm$ 2.2} & \multirow{2}{*}{$-$90.0 $\pm$ 1.7} \\
     {\NI} & 1199.6 & 215.6 $\pm$ 23.7 &  &  \\
     {\NI} & 1200.2 & 212.6 $\pm$ 25.6 &  &  \\
     {\NI} & 1200.7 & 189.5 $\pm$ 25.1 &  &  \\
     \hline
     {\MgII} & 2796.4 & 518.1 $\pm$ 142.1 & 13.38 $\pm$ 0.50 & 24.8 $\pm$ 10.3 & $-$100.0 $\pm$ 24.1 \\
     {\MgII} & 2803.5 & $<$511.8 &  &  \\
     \hline
     {\SiII} & 1260.4 & 567.5 $\pm$ 21.4 & \multirow{3}{*}{14.36 $\pm$ 0.06} & \multirow{3}{*}{47.9 $\pm$ 2.5} & \multirow{3}{*}{$-$94.9 $\pm$ 2.0} \\
     {\SiII} & 1193.3 & 491.0 $\pm$ 34.5 &  &  \\
     {\SiII} & 1190.4 & 399.1 $\pm$ 35.6 &  &  \\
     \hline
     {\SII} & 1259.5 & $<$57.6 & $<$14.40 & ... & ... \\
     {\SII} & 1253.8 & $<$66.6 & & \\
     {\SII} & 1250.6 & $<$63.3 & & \\
     \hline
     {\CII} & 1036.3 & 541.6 $\pm$ 38.7 & 15.56 $\pm$ 0.56 & 52.6 $\pm$ 15.4 & $-$96.1 $\pm$ 5.3\\
     \hline
     {\NII} & 1083.0 & 400.1 $\pm$ 19.3 & 15.13 $\pm$ 0.12 & 43.6 $\pm$ 5.2 & $-$91.9 $\pm$ 2.7 \\
     \hline
     {\SiIII} & 1206.5 & 470.5 $\pm$ 37.8 & 13.91 $\pm$ 0.17 & 49.1 $\pm$ 8.0 & $-$96.2 $\pm$ 4.2 \\
      \hline
     {\CIV} & 1548.2 & 497.7 $\pm$ 123.0 & 14.23 $\pm$ 0.12 & 124.8 $\pm$ 1.4 & -88.4 $\pm$ 28.6 \\
     {\CIV} & 1550.8 & ...$^{d}$ & \\ 
     \hline
     {\NV} & 1238.8 & $<$72.3 & \\ 
     {\NV} & 1242.8 & 105.9 $\pm$ 32.2 & 14.13 $\pm$ 0.13 & 72.9 $\pm$ 28.4 & $-$105.0 $\pm$ 19.8 \\
     \hline
     {\OVI} & 1031.9 & 324.0 $\pm$ 55.0 & 14.62 $\pm$ 0.09 & 81.7 $\pm$ 20.6 & $-$77.9 $\pm$ 15.2 \\
     {\OVI} & 1037.6 & ...$^{d}$ & \\ 
     \enddata
    \tablecomments{ $^a$ Transitions not detected at 3$\sigma$ above the noise level are reported as the 3$\sigma$ upper limit. $^b$ Transitions which are not detected are assumed to be in the linear portion of the curve of growth to convert $W_r$ to $N$. $^c$ Doppler widths of each level of {\mH} ($J$) were first fit independently of the others, then were refit while fixing the Doppler width to be the weighted mean, as described in the text. $^d$ A measurement could not be made due to the line being blended with an intervening system.}
\end{deluxetable}

\bibliography{References}
\bibliographystyle{aasjournalv7.1}

\end{document}